\documentclass[aps,prl,amsfonts,amssymb,twocolumn,amsmath,preprintnumbers,nofootinbib,floatfix,superscriptaddress]{revtex4-2}
\usepackage[colorlinks=true, letterpaper=true, pdfstartview=FitV, linkcolor=blue, citecolor=blue, urlcolor=blue]{hyperref}
\usepackage{amsfonts}
\usepackage{mathrsfs}
\usepackage{amsmath}% needed for subequations
\usepackage{color}
\usepackage{physics}
\usepackage{siunitx} % si unit
\usepackage{graphicx}
\usepackage{bm}% bold maths
\usepackage{amssymb}
\usepackage{xspace}
\usepackage{epstopdf}
\usepackage{multirow}
\usepackage{dcolumn}% Align table columns on decimal point
\usepackage{longtable}
\usepackage{multirow}
\usepackage{float}
\usepackage{comment}
\usepackage[normalem]{ulem}
\usepackage{braket}

\DeclareDocumentCommand\vectorbold{ s m }{\IfBooleanTF{#1}{\boldsymbol{#2}}{\mathbf{#2}}}

\makeatother
\begin{document}

\title{Quantum Geometric Origin of Nonlinear Current Induced Orbital Magnetization}

\author{Xue-Jin Zhang}
\thanks{These authors contributed equally to this work.}
\affiliation{Institute of Applied Physics and Materials Engineering, University of Macau, Macau, China}

\author{Yue-Xin Huang}
\email{hyx@gbu.edu.cn}
\thanks{These authors contributed equally to this work.}
\affiliation{School of Sciences, Great Bay University, Dongguan 523000, China}
\affiliation{Great Bay Institute for Advanced Study, Dongguan 523000, China}

\author{Wei Du}
\thanks{These authors contributed equally to this work.}
\affiliation{State Key Laboratory of Surface Physics and Interdisciplinary Center for Theoretical Physics and Information Sciences, Fudan University, Shanghai 200433, China}

\author{Xiaolong Feng}
\thanks{These authors contributed equally to this work.}
\affiliation{Max Planck Institute for Chemical Physics of Solids, Nöthnitzer Strasse 40, D-01187 Dresden, Germany}

\author{Shen Lai}
\affiliation{Institute of Applied Physics and Materials Engineering, University of Macau, Macau, China}

\author{Cong Xiao}
\email{congxiao@fudan.edu.cn}
\affiliation{State Key Laboratory of Surface Physics and Interdisciplinary Center for Theoretical Physics and Information Sciences, Fudan University, Shanghai 200433, China}

\author{Qian Niu}
\affiliation{CAS Key Laboratory of Strongly-Coupled Quantum Matter Physics, and Department of Physics,
University of Science and Technology of China, Hefei, Anhui 230026, China}

\author{Shengyuan A. Yang}
\affiliation{Research Laboratory for Quantum Materials, Department of Applied Physics, The Hong Kong Polytechnic University, Hong Kong, China}

\begin{abstract}
Electric generation of magnetization is a focus of condensed matter research, and has recently been advanced into the nonlinear regime. However, due to the nonlocal nature of orbital magnetism, how to properly formulate nonlinear current-induced orbital magnetization remains a fundamental challenge. Here, we develop the proper theory for this effect. This is based on the microscopic derivation of field-corrected orbital magnetic moment of a Bloch electron,
a critical missing piece in the present theory. We show that the quantum geometric origin of this phenomenon lies in both the anomalous orbital polarizability and the Berry-connection polarizability, which often provide competing contributions. Combining our theory with
first-principles calculations, we predict significant, experimentally accessible nonlinear orbital magnetization generated in strained bilayer graphene, monolayer 1T' $\mathrm{MoS_2}$ and $\mathrm{MoTe_2}$. Remarkably, nonlinear orbital magnetization can dominate over its spin counterpart in materials with topological band features, irrespective of the spin-orbit coupling strength.
\end{abstract}
\maketitle

Electric generation of magnetization has been a problem of long-lasting interest in condensed matter physics.
In \emph{nonmagnetic} crystals, this generation is from the magnetic moments of non-equilibrium carriers driven by applied electric current~\cite{pikus1978,aronov1989,edelstein,culcer2007,Awschalom2009,Murakami2015,Zhong2016,Pesin2017,Murakami2018orbital,kato2004,Stern2006,Mak2017valley,furukawa2017observation}. At linear order in the current (or $E$ field), the effect is forbidden in crystals with inversion symmetry.
Recently, the possibility of nonlinear generation, with induced magnetization $\delta M\propto E^2$, has been proposed and attracted great attention~\cite{Xiao2022NLSOT,Wang2022,xiao2023,Kodama2024,Guo2024,oike_2024,oike_2024a,Feng2025Quantum,Wang_2025NLSG,wang_2026NSP,baek_NEdelstein_2024,Liao2024nonlinear-OM,Liao2025nonlinear-OM,Wang2026SMOKE,Lu2026Christoffel}. Such nonlinear response can dominate in centrosymmetric crystals, and more importantly,
it may manifest quantum geometric properties of Bloch electrons not accessible in linear order response effects.
For example, Ref.~\cite{xiao2023} developed the theory of nonlinear current-induced spin magnetization, which originates from
the momentum-space dipole of anomalous spin polarizability (ASP).
The geometric quantity ASP describes the induced anomalous spin moment of a spin-orbit coupled Bloch electron in an electric field~\cite{xiao2023}, in analogy to the famous anomalous velocity induced by Berry curvature~\cite{chang_berry_1995,jungwirth2002,xiao2010}.

Besides spin contribution, magnetization in general also has an orbital contribution. At linear order in $E$, this current-induced orbital magnetization (CIOM) is obtained as~\cite{Murakami2015,Zhong2016,Mak2017valley,furukawa2017observation}
\begin{eqnarray}\label{CIOM}
\delta \bm M= \sum_\ell \delta f_\ell \bm m_\ell,
\end{eqnarray}
where $\ell$ labels the electronic states, $\delta f_\ell\propto E$ is the off-equilibrium distribution, and $\bm m_\ell$ is the orbital magnetic moment of a Bloch electron~\cite{Chang1996,sundaram_wave-packet_1999,xiao2010}. It is crucial to note that, unlike spin, due to the nonlocal nature of orbital magnetic dipole operator, $\bm m_\ell$ \emph{cannot} be directly evaluated in the extended Bloch basis; instead, it has to be rigorously derived for a Bloch wave packet state, as well developed in Refs.~\cite{Chang1996,sundaram_wave-packet_1999,xiao2010}.

Now, to study the effect of nonlinear CIOM, evidently, one has to know $\delta \bm m_\ell$, i.e., the first-order $E$ field correction to the orbital magnetic moment. However, this critical piece of physics has not been properly formulated yet.
In fact, this difficulty is the reason underlying contradicting results reported in several recent works on nonlinear CIOM~\cite{Liao2024nonlinear-OM,Liao2025nonlinear-OM,Wang2026SMOKE,Lu2026Christoffel}, where the form of $\delta \bm m_\ell$ was speculated or phenomenologically argued, without a rigorous justification.
Therefore, it remains an outstanding challenge to formulate the field correction to orbital moment $\delta \bm  m_\ell$, to establish the proper theory of nonlinear CIOM, and further, to unveil the quantum geometric origin of this effect.

We solve this challenge in this work. We develop the proper theory for field-corrected orbital magnetic moments of Bloch electrons. Based on this fundamental result, we formulate the theory of nonlinear CIOM in nonmagnetic crystals. We show that this effect originates from two basic geometric properties of Bloch band structure: anomalous orbital polarizability (AOP)~\cite{gao2015,Wang2024IPHE} and Berry-connection polarizability (BCP)~\cite{gao2014,liu2022third}. From this understanding, we predict that nonlinear CIOM can dominate over its spin counterpart in multiband structures with topological band features, regardless of the spin-orbit coupling (SOC) strength.
Combining our theory with first-principles calculations, we report significant nonlinear orbital magnetization that can be detected in three representative materials: strained bilayer graphene with negligible SOC, 1T' monolayer $\mathrm{MoS_2}$ with moderate SOC, and 1T' monolayer $\mathrm{MoTe_2}$ with strong SOC. In all these examples, the induced orbital magnetization is indeed found to be dominating, at least two orders of magnitude greater than its spin counterpart.
These findings fill a critical missing piece in the theory of Bloch electrons, establish a solid foundation for investigating magnetoelectric effects, and offer guidance for exploring significant CIOM in experimentally accessible materials.

%timely clarify a controversial issue in fundamental electromagnetism (the $E$-field induced anomalous orbital moment for each electron), lay the framework for studying nonlinear electrical generation of orbital magnetization, unveil the significance of CIOM in broad platforms beyond weak-SOC systems, and predict detectable effects in experimentally accessible materials.

\emph{\color{blue}Field correction of orbital magnetic moment.} The key missing piece in the present theory is how the orbital magnetic moment of a Bloch electron is corrected by an applied $E$ field.

First, recall that in the absence of $E$ field, the orbital magnetic moment is evaluated from the expectation value of the magnetic dipole~\cite{Chang1996,sundaram_wave-packet_1999,xiao2010}:
\begin{eqnarray}\label{lm}
  \bm m_{n\bm k}=-\frac{e}{2}\text{Re} \braket{W_{n\bm k}|(\hat{\bm r}-\bm r_c)\times \hat{\bm v}|W_{n\bm k}},
\end{eqnarray}
where a Bloch electron is described by a Bloch wave packet $\ket{W_{n\bm k}}$ centered at Bloch state $\ket{\psi_{n\bm k}}$, $n$ is the band index, $\bm k$ is the Bloch wave vector,  $(-e)$ is the electron charge, $\bm r_c\equiv \braket{W|\hat{\bm r}|W}$ is the central position of the wave packet, and $\hat{\bm v}$ is the velocity operator.
Note that because of the presence of position operator $\hat{\bm r}$, Eq.~(\ref{lm}) cannot be simply reduced to some average at the central Bloch state $\ket{\psi_{n\bm k}}$. In fact, the position operator is ill-defined in the Bloch basis (which are extended states). This is in contrast to spin moment (and other quantities), whose operator is local and well-defined in Bloch basis, making such a reduction legitimate. However, for $\bm m_{n\bm k}$, one has to explicitly deal with the wave packet state, and the evaluation gives the well-known result~\cite{Chang1996,sundaram_wave-packet_1999,xiao2010}:
\begin{eqnarray}
  \bm m_{n\bm k}=-\frac{e}{2}\text{Re}\sum_{n'\neq n}\bm{\mathcal{A}}_{nn'}\times \bm{v}_{n'n},
\end{eqnarray}
where $\bm{v}_{n'n}$ is the interband velocity matrix element, and $\bm{\mathcal{A}}_{nn'}=\braket{u_{n\bm k}|i\nabla_{\bm k}u_{n'\bm k}}$ is the interband Berry connection, with $\ket{u}$ being the cell-periodic part of Bloch state $\ket{\psi}$.

An applied $E$ field perturbs the wave packet hence corrects the orbital moment.
The corrected moment is given by
\begin{eqnarray}\label{mE}
  \tilde{\bm m}_{n\bm k}=-\frac{e}{2}\text{Re} \braket{\tilde W_{n\bm k}|(\hat{\bm r}-\tilde{\bm r}_c)\times \hat{\bm v}|\tilde W_{n\bm k}},
\end{eqnarray}
where we use tilde to indicate quantities including $E$ field correction. It is important to note that the wave packet center $\tilde{\bm r}_c$ also receives a correction~\cite{gao2014}, since it is determined from the perturbed state $\ket{\tilde W}$.

To obtain the first-order correction $\delta \bm m_{n\bm k}\propto E$, one expands all tilde quantities in Eq.~(\ref{mE})
in powers of $E$ and collects the first-order terms. The result comprises two parts: $\delta \bm m=\delta \bm m^\text{I}+\delta \bm m^\text{II}$. $\delta \bm m^\text{I}$ is from the correction in $\tilde{\bm r}_c$ (to simplify notation, we will drop the state labels $n$ and $\bm k$ when the context is clear):
\begin{eqnarray}
  \delta {\bm m}^\text{I}=\frac{e}{2}\text{Re} \braket{ W|\bm{\mathfrak a}^E\times \hat{\bm v}|W},
\end{eqnarray}
where $\mathfrak a^E_a=\mathcal G_{ab} E_b$ is the field-induced position shift of the wave packet center~\cite{gao2014} ($a, b$ are Cartesian indices, and repeated indices are summed over), and $\mathcal G_{ab}$ is the BCP given by~\cite{gao2014,liu2022third,wang2021,liu2021,Xiao2025definition}
\begin{eqnarray}
    \mathcal G_{ab}=2e\, \mathrm{Re}\sum_{n'\ne n}
    \frac{(\mathcal A_a)_{n n'} (\mathcal A_b)_{n' n}}{\varepsilon_n-\varepsilon_{n'}},
    \label{eq-G}
\end{eqnarray}
with $\varepsilon_n$ being the band energy of $\ket{u_n}$. Meanwhile, $\delta \bm m^\text{II}$ is from the correction $\ket{\delta W}$ of wave packet in Eq.~(\ref{mE}):
\begin{eqnarray}
     \delta \bm m^\text{II}=-\frac{e}{2}\text{Re}
    \braket{ W| [(
{\hat{\bm r}}-\bm{r}_{c}) \times {\hat{\bm v}}- {\hat{\bm v}} \times (
{\hat{\bm r}}-\bm{r}_{c})] |\delta W}.
\end{eqnarray}
The evaluation of $\delta \bm m^\text{II}$ is a bit complicated (see Supplemental Material~\cite{supp}).
After calculation, we obtain
\begin{eqnarray}
    \delta \bm m^\text{I} & = & \frac{e}{2}\bm{\mathfrak a}^{E}\times\bm{v}, \\
    \delta \bm m^\text{II} & = &\frac{e}{2}\bm{\mathfrak a}^{E}\times\bm{v} -eE_a \mathcal F_{ab}\hat{\bm e}_b,
    \label{AOP}
\end{eqnarray}
where $\bm v$ is the band velocity, and $\hat{\bm e}_b$ is the unit basis vector along $b$. $\mathcal{F}_{ab}$ is the AOP tensor given by~\cite{gao2015,Wang2024IPHE,Wang2024IPHE-exp}
\begin{eqnarray}\label{dAOP}
    \mathcal{F}_{ab}= -
2\mathrm{{\operatorname{Re}}}\sum_{n'\neq n}\frac{(\mathcal{A}_a)_{nn'}(\mathcal{M}_b)_{n'n}}
{\varepsilon_{n}-\varepsilon_{n'}}
+\frac{e}{2\hbar}\epsilon_{ijb}\Gamma_{ija},
\end{eqnarray}
where $\boldsymbol{\mathcal M}_{n'n}=e\sum_{l\ne n}(\bm v_{n' l}+\delta_{ln'}\bm v_n)\times \boldsymbol{\mathcal A}_{ln}/2$
mimics an interband orbital moment, $\epsilon_{ijb}$ is the Levi-Civita symbol,
$
    \Gamma_{ija}=\frac{1}{2}(-\partial_i g_{ja}+\partial_j g_{ai}+\partial_a g_{ji})
$
is the Christoffel symbol~\cite{gao2014} associated with the $k$-space quantum metric $g_{ab}=\mathrm{{\operatorname{Re}}}\sum_{n'\neq n}
(\mathcal{A}_a)_{nn'}(\mathcal{A}_b)_{n'n}$, and $\partial_a\equiv \partial_{k_a}$. The quantum metric measures distance in $k$-space. The appearance of such a metric-related term is a manifestation of the nonlocal (in $k$-space) nature of orbital moment.

\begin{figure*}[t!]
    \centering
    \includegraphics[width=\textwidth]{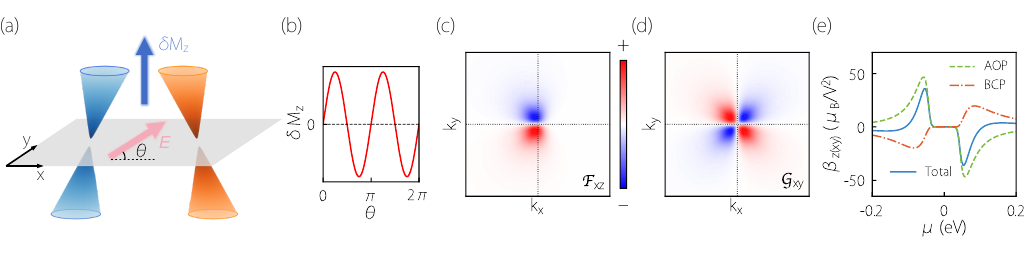}
    \caption{(a) Schematic spectrum of the 2D Dirac model (\ref{eq-WeylModel}), where an in-plane $E$ field generates a nonlinear out-of-plane $\delta M_z$. (b) Angular dependence of $\delta M_z$. (c-d) $k$-space distribution of AOP $\mathcal{F}^x_z(\bm k)$ and BCP $\mathcal G_{xy}(\bm k)$ of the conduction band at $\zeta=+$ valley. (e) Calculated nonlinear CIOM response coefficient $\alpha_{z(xy)}$ versus the chemical potential $\mu$. In the calculation, we take $v_F=\SI{1}{eV\cdot\angstrom}$, $w=0.6v_F$, $\Delta=\SI{50}{meV}$, and $\tau=\SI{50}{fs}$.}
    \label{fig-fig1}
\end{figure*}

Collecting terms, we can obtain the $E$-field correction of orbital magnetic moment of a Bloch electron as
\begin{eqnarray}\label{dme}
    \delta \bm m_{n\bm k}= -eE_a \mathcal{F}_{ab} \hat{\bm e}_b
    +e (\mathcal G_{ab}E_b\hat{\bm{{e}}}_a)\times \bm v.
\end{eqnarray}
This is the first key result of this paper.
We have a few remarks. First, it is known that AOP characterizes the positional shift of wave packet center induced by an orbital magnetic field \cite{gao2014}: $\mathfrak a^{B}_a=\mathcal{F}_{ab} B_b$.
It hence leads to a magnetoelectric coupling energy $\sim \bm E\cdot \bm{\mathfrak a}^{B}\propto EB$ for the wave packet. The AOP contribution in (\ref{dme}) can be understood as a result of this energy. Second, the BCP contribution to $\delta\bm m$ may be intuitively understood as the $E$-field induced variation of self-rotating radius of the wave packet.
Third, previous studies on nonlinear transport effects have revealed the significance of BCP and AOP, where they separately contribute to different types of transport \cite{gao2014,wang2021,liu2021,liu2022third,Gao2023QM,Han2024room,Xiao2025definition,Wang2024IPHE,Wang2024IPHE-exp}. Here gives a scenario where they make a combined contribution to
the electrically induced orbital moment.

\emph{\color{blue} Theory of nonlinear CIOM.}
The orbital magnetization generated at the second order of $E$ field can be expressed as
\begin{eqnarray}
    \delta M_a= \beta_{abc} E_b E_c,
    \label{symmetry}
\end{eqnarray}
where $\beta_{abc}$ is the pertaining response tensor.
We are considering nonmagnetic crystals, so $\beta_{abc}$ is a time-reversal ($\mathcal T$) even rank-3 pseudotensor. Its symmetry property can be easily analyzed (see Supplemental Material \cite{supp}). Particularly, we note that among the 14 non-gyrotropic point groups, where linear CIOM is prohibited, 12 of them allow the nonlinear CIOM (listed in \cite{supp}). In other words, in these 12 crystal classes, nonlinear CIOM would be the leading order contribution to orbital magnetoelectric response.

Next, since magnetization and $E$ field have opposite parity under $\mathcal T$, the $\mathcal T$-even character of the response requires carrier relaxation process around Fermi surface, with off-equilibrium distribution
$
    \delta f \propto \tau E
$ in Eq.~(\ref{CIOM}), where $\tau$ is the relaxation time.
Rewriting Eq.~(\ref{CIOM}) using Bloch state labels and retaining the contribution $\propto E^2$, the nonlinear orbital magnetization is given by
\begin{eqnarray}
    \delta \bm M= \int[\dd{\bm k}] \delta f_{n \bm k} \delta {\bm m}_{n\bm k},
    \label{kinematic M}
\end{eqnarray}
where $[\dd{\bm k}]$ denotes $\sum_n \dd{\bm k}/(2\pi)^d$, with $d$ being the dimension of the system.
Substituting Eq.~(\ref{dme}) and
$
    \delta f = e \tau E_a\partial_a f_0/\hbar
$ ($f_0$ is the equilibrium distribution) into (\ref{kinematic M}), one obtains the nonlinear CIOM response tensor as
\begin{eqnarray}\label{beta}
  \beta_{abc}=\tau\frac{e^2}{\hbar}\int[\dd{\bm k}] f_0
    \partial_c \left( \mathcal{F}_{ba} + \epsilon_{ija} v_i \mathcal G_{jb} \right).
\end{eqnarray}
This formula is another key result of this work.

One observes that both AOP and BCP contribute to nonlinear CIOM. Actually, the integral in Eq.~(\ref{beta}) can be regarded as a momentum space dipole of their combination $( \mathcal{F}_{ba} + \epsilon_{ija} v_i \mathcal G_{jb})$.
Below, we shall see that contributions from AOP and BCP are generally comparable in $\beta$.

\emph{\color{blue}A model study.}
To illustrate the feature of nonlinear CIOM, we first apply our theory to a two-dimensional (2D) massive Dirac model. Consider two Dirac valleys labeled with valley index $\zeta=\pm$, connected by $\mathcal T$ and $\mathcal{M}_x$ symmetries. Their Hamiltonian reads
\begin{eqnarray}
    H_\zeta= \zeta w k_x +v_F(\zeta k_x \sigma_x+k_y \sigma_y) + \Delta \sigma_z,
    \label{eq-WeylModel}
\end{eqnarray}
where $\sigma$'s are Pauli matrices denoting an orbital degree of freedom, $w$, $v_F$, and $\Delta$ are real model parameters. Figure~\ref{fig-fig1}(a) shows the typical band structure of this model.
%The two Weyl points are connected by $\mathcal T=\sigma_y \mathcal K$ and $\mathrm{M_x}=\sigma_y$, namely, the model preserve time-reversal symmetry and mirror reflection along $x$.

In this 2D model, due to $\mathcal T$ and $\mathcal{M}_x$ symmetries, only $\beta_{z(xy)}\equiv(\beta_{zxy}+\beta_{zyx})/2$ is symmetry allowed,
%(And to have a nonzero $\beta_{z(xy)}$, the tilt ($w$) term in (\ref{eq-WeylModel}) is needed, to break the isotropy of a valley.)
and the response takes the form of
\begin{eqnarray}
    \delta M_z= \beta_{z(xy)}E^2\sin 2\theta,\label{symmetry}
\end{eqnarray}
where
$\theta$ is the angle between the in-plane $E$ field and the $x$ axis. The response has a $\pi$ period, and vanishes when $E$ is perpendicular or parallel to the  $\mathcal{M}_x$ mirror line (see Fig.~\ref{fig-fig1}(b)).

In Figs.~\ref{fig-fig1}(c,d), we plot AOP and BCP for the conduction band at $\zeta=+$ valley (results for the other valley are the same, as required by $\mathcal{T}$ symmetry).
{One sees that $\mathcal G_{xy}$ displays a quadrupole-like pattern, whereas $\mathcal{F}_{xz}$ shows a skewed dipole-like pattern.}
More importantly, both quantities have their peak values concentrated around band edge, which can be expected since as geometric quantities, they manifest interband coherence. As such, the nonlinear CIOM should be enhanced near small-gap regions in a band structure.

This expectation is confirmed in Fig.~\ref{fig-fig1}(e), which shows the result of $\beta_{z(xy)}$ as a function of chemical potential $\mu$. We also separately plot
AOP and BCP contributions. One finds that (i) all these contributions are peaked near band edges; (ii) they have a sign change crossing the gap; and (iii) AOP and BCP contributions have comparable magnitudes. Interestingly, they exhibit opposite signs in this model, leading to a strong competition. In Fig.~\ref{fig-fig1}(e), one sees that AOP dominates the competition near band edge, but BCP wins at higher $\mu$.

\emph{\color{blue} Orbital versus spin responses in real materials.}
CIOM does not need SOC~\cite{Zhong2016,Murakami2018orbital}, but its spin counterpart must require SOC.
This implies immediately that CIOM dominates over spin response in weak-SOC materials. For example, we find that in strained bilayer graphene,
where SOC (hence the spin response) is negligible, nonlinear CIOM can achieve a significant magnitude $\sim \SI{2.5e4}{\mu_B/V^2}$~\cite{supp}.

Moreover, we stress that the dominance of nonlinear CIOM can be expected even in strong-SOC materials, provided that Fermi level is located around band near-degeneracies. From the model study above, we have seen both AOP and BCP are enhanced around
small-gap regions. In fact, from their expressions, one can see that they scale as $1/\Delta^3$, with $\Delta$ being the local direct gap. In comparison, the ASP for nonlinear spin response scales as $1/\Delta^2$~\cite{xiao2023}. This scaling behavior implies the orbital response tends to dominate the spin response when $\Delta$ is small.

\begin{figure}[!t]
    \centering
    \includegraphics[width=0.48\textwidth]{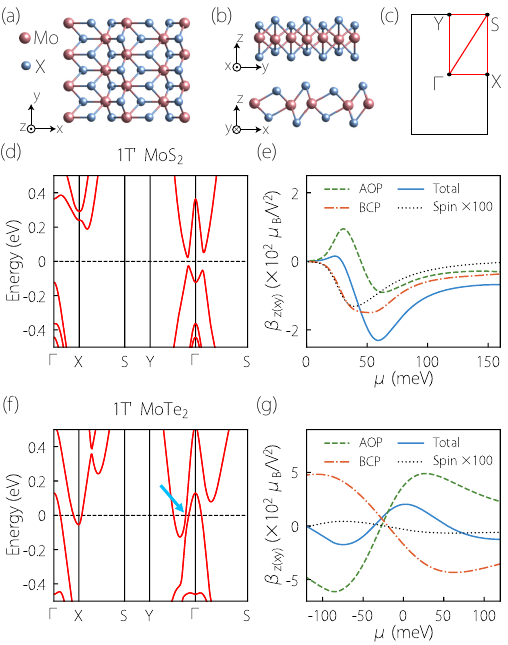}
    \caption{(a) Top and (b) side views of monolayer 1T' $\mathrm{MoX_2}$ ($\mathrm{X}=\mathrm{S}$ or $\mathrm{Te}$). (c) Brillouin zone. (d) Calculated
    band structures and (e) nonlinear CIOM coefficient $\beta_{z(xy)}$ for 1T' $\mathrm{MoS_2}$.
    (f) and (g) show the corresponding results for 1T' $\mathrm{MoTe_2}$. In (e) and (g), the results for nonlinear current-induced spin magnetization are also shown with dotted lines (multiplied by a factor of 100).
    In the calculation, we set $T=\SI{50}{K}$  and $\tau=50$ fs.}
    \label{fig-MoX2}
\end{figure}

To demonstrate the above rationale, we combine our theory with first-principles DFT calculation to
perform a comparative study on monolayer 1T' $\mathrm{MoX}_2$ ($\text{X}=$\ S, Te), which have been experimentally realized and extensively studied in the past decade~\cite{qian2014,wu_Observation_2018,katsuragawa_Roomtemperature_2020,naylor2016,tang2018,peng_High_2019}. As shown in Figs.~\ref{fig-MoX2}(a,b), in their 2D structure, the three atomic layers adopt a rhombohedral stacking, belonging to space group $P2_1/m$ (No.~$11$) and point group $\mathrm{C_{2h}}$.
The linear CIOM is symmetry-prohibited, whereas the out-of-plane nonlinear CIOM is allowed and takes the form of Eq.~(\ref{symmetry}).

Figure~\ref{fig-MoX2}(d) shows the calculated band structures of 1T' $\mathrm{MoS_2}$. It is a
small-gap semiconductor, with a band gap around $\Gamma$ point opened by its moderate SOC, consistent with previous works~\cite{qian2014}.
Figure~\ref{fig-MoX2}(e) presents the calculated $\beta_{z(xy)}$ for $\mathrm{MoS_2}$ as well as its AOP and BCP components as a function of $\mu$ in conduction band. Here, we take $\tau=\SI{50}{fs}$, which is typical for 2D materials at low temperatures~\cite{ma2019}.
Like in the model study, the AOP and BCP terms exhibit comparable magnitudes and are both pronounced when $\mu$ is located near the conduction band bottom. Due to their competition, the total $\beta_{z(xy)}$ displays a complex energy dependence, being non-monotonic and undergoing sign changes with $\mu$.
Here, the maximal orbital response reaches a magnitude $\sim\SI{2.3e2}{\mu_B/V^2}$.
For comparison, we also estimate the nonlinear spin magnetization according to Ref.~\cite{xiao2023}. As shown in Fig.~\ref{fig-MoX2}(e), the nonlinear spin response is two orders of magnitude smaller than the orbital response.

$\mathrm{MoTe_2}$ is isostructural with $\mathrm{MoS_2}$, but has a stronger SOC.
Indeed, the strong SOC significantly changes the low-energy band structure and turns 1T' $\mathrm{MoTe_2}$ into a semimetal (see Fig.~\ref{fig-MoX2}(f)).
Will the spin response be more important or even outweigh the orbital one?
The answer is negative. As shown in Fig.~\ref{fig-MoX2}(g), the orbital response remains dominating over spin by more than two orders of magnitude. The total $\beta_{z(xy)}$ reaches a large value $\sim \SI{2e2}{\mu_B/V^2}$ at intrinsic Fermi level.
Here, the AOP and BCP are enhanced by the band near-degeneracy region as indicated in Fig.~\ref{fig-MoX2}(f).
Again, they possess comparable magnitudes, and exhibit opposite signs in a range of energy, similar to  Fig.~\ref{fig-fig1}(e).

Under a moderate $E$ field of $10^5$ V/m~\cite{jungwirth2018,Xiao2022NLSOT}, the induced nonlinear orbital magnetization can reach $\sim \SI{2e-6}{\mu_B/nm^2}$ in these 2D monolayers. This value is quite significant and is larger than previously reported values ($\sim 10^{-7}$ to $\SI{e-6}{\mu_B/nm^2}$) by linear magnetoelectric response in non-centrosymmetric systems~\cite{kato2004,Stern2006,Mak2017valley}, indicating that the nonlinear CIOM signal should be readily detected in experiment.

\emph{\color{blue}Discussion.}
The field correction of orbital magnetic moment [Eq.~(\ref{dme})] is a fundamental missing piece in the theory of Bloch electrons. Based on this result, we formulate the theory for nonlinear CIOM [Eq.~(\ref{beta})] and reveal its origin in AOP and BCP, which are gauge-invariant quantum geometric properties of Bloch electrons.
This helps to resolve the controversy among existing theories, which arises from the incomplete formulation of $\delta \bm m$. 
As we demonstrated, AOP and BCP contributions are generally comparable; missing either one of them may completely change the physical result. In addition, our theory also leads to an important prediction that nonlinear CIOM should outperform its spin counterpart when the material's band structure has near-degeneracies near Fermi level, due to the enhancement in AOP and BCP. This phenomenon can occur even in materials with strong SOC, as demonstrated here using concrete materials.

Beyond centrosymmetric systems, the effect may also play an important role in inversion breaking materials. Particularly, in $2$D systems, with in-plane driving field, any out-of-plane rotation axis forbids the out-of-plane magnetization in linear response, but nonlinear generation is permitted. For instance,
non-centrosymmetric 2D materials with $\mathrm{C_{3h}}$ group and chiral polar point groups $\mathrm{C_{n}}$ ($n=2,3,4,6$) can support a nonlinear out-of-plane magnetization induced by in-plane $E$ field. In addition, nonlinear orbital magnetization can also be the leading effect on the 2D surface of certain 3D materials. For example, in a 3D material with bulk dihedral point group symmetries $\mathrm{D_{n}}$ ($n=2,3,4,6$), the nonlinear magnetization along $z$ (along principal axis) is forbidden in bulk, but is allowed at the top and bottom surfaces.

The proposed effect may find useful application in spintronics. The CIOM in a nonmagnetic layer can be used to drive magnetic dynamics in a neighboring magnetic layer. If this bilayer structure has some out-of-plane rotation symmetry,
then linear CIOM can only produce an in-plane magnetization, whereas nonlinear CIOM may generate an unconventional out-of-plane magnetization. Such an out-of-plane component may play a critical role, e.g., in magnetic switching processes~\cite{lee2026orbital,Klaui2026orbital}, which is an interesting topic for future studies.

\bibliography{ref}

@misc{supp,
  howpublished = {See Supplemental Material for computational details.}
}

@article{Xiao2025definition,
  title     = {Proper Definition of Intrinsic Nonlinear Current},
  author    = {Xiao, Cong and Cao, Jin and Niu, Qian and Yang, Shengyuan A.},
  journal   = {Phys. Rev. Lett.},
  volume    = {135},
  issue     = {25},
  pages     = {256306},
  numpages  = {9},
  year      = {2025},
  month     = {Dec},
  publisher = {American Physical Society},
  doi       = {10.1103/g6yz-pwzh},
  url       = {https://link.aps.org/doi/10.1103/g6yz-pwzh}
}

@article{Chang1996,
  title     = {Berry phase, hyperorbits, and the Hofstadter spectrum: Semiclassical dynamics in magnetic Bloch bands},
  author    = {Chang, Ming-Che and Niu, Qian},
  journal   = {Phys. Rev. B},
  volume    = {53},
  issue     = {11},
  pages     = {7010--7023},
  numpages  = {0},
  year      = {1996},
  month     = {Mar},
  publisher = {American Physical Society},
  doi       = {10.1103/PhysRevB.53.7010},
  url       = {https://link.aps.org/doi/10.1103/PhysRevB.53.7010}
}

@article{baek_NEdelstein_2024,
  title   = {Nonlinear Orbital and Spin {{Edelstein}} Effect in Centrosymmetric Metals},
  author  = {Baek, Insu and Han, Seungyun and Cheon, Suik and Lee, Hyun-Woo},
  year    = 2024,
  month   = jul,
  journal = {npj Spintron.},
  volume  = {2},
  number  = {1},
  pages   = {33},
  issn    = {2948-2119},
  doi     = {10.1038/s44306-024-00041-4},
  url     = {https://www.nature.com/articles/s44306-024-00041-4}
}

@article{oike_2024,
  title   = {Impact of Electron Correlations on the Nonlinear {{Edelstein}} Effect},
  author  = {{\=O}ik{\'e}, Jun and Peters, Robert},
  year    = 2024,
  month   = oct,
  journal = {Phys. Rev. B},
  volume  = {110},
  number  = {16},
  pages   = {165111},
  issn    = {2469-9950, 2469-9969},
  doi     = {10.1103/PhysRevB.110.165111},
  url     = {https://link.aps.org/doi/10.1103/PhysRevB.110.165111}
}

@article{Wang_2025NLSG,
  title   = {Quantum-Metric-Induced Plateau of Nonlinear Spin Generation},
  author  = {Lu, G. T. and Ma, Yuxuan and Wang, C. M.},
  year    = 2025,
  month   = may,
  journal = {Phys. Rev. B},
  volume  = {111},
  number  = {19},
  pages   = {195428},
  issn    = {2469-9950, 2469-9969},
  doi     = {10.1103/PhysRevB.111.195428},
  url     = {https://link.aps.org/doi/10.1103/PhysRevB.111.195428}
}

@article{wang_2026NSP,
  title   = {Current-Induced Time-Reversal-Even Nonlinear Spin Polarization in p -Wave Magnets},
  author  = {Niu, Ren-Zhao and Duan, Hou-Jian and Ma, Rong and Deng, Ming-Xun and Wang, Rui-Qiang},
  year    = 2026,
  month   = jan,
  journal = {Phys. Rev. B},
  volume  = {113},
  number  = {4},
  pages   = {045101},
  issn    = {2469-9950, 2469-9969},
  doi     = {10.1103/j73g-flft},
  url     = {https://link.aps.org/doi/10.1103/j73g-flft}
}

@article{oike_2024a,
  title   = {Nonlinear Magnetoelectric Effect under Magnetic Octupole Order: {{Application}} to a d -Wave Altermagnet and a Pyrochlore Lattice with All-in/All-out Magnetic Order},
  author  = {{\=O}ik{\'e}, Jun and Shinada, Koki and Peters, Robert},
  year    = 2024,
  month   = nov,
  journal = {Phys. Rev. B},
  volume  = {110},
  number  = {18},
  pages   = {184407},
  issn    = {2469-9950, 2469-9969},
  doi     = {10.1103/PhysRevB.110.184407},
  url     = {https://link.aps.org/doi/10.1103/PhysRevB.110.184407}
}

@article{Pesin2017,
  title     = {Kinetic orbital moments and nonlocal transport in disordered metals with nontrivial band geometry},
  author    = {Rou, J. and \ifmmode \mbox{\c{S}}\else \c{S}\fi{}ahin, C. and Ma, J. and Pesin, D. A.},
  journal   = {Phys. Rev. B},
  volume    = {96},
  issue     = {3},
  pages     = {035120},
  numpages  = {12},
  year      = {2017},
  month     = {Jul},
  publisher = {American Physical Society},
  doi       = {10.1103/PhysRevB.96.035120},
  url       = {https://link.aps.org/doi/10.1103/PhysRevB.96.035120}
}

@article{Lu2026Christoffel,
  title     = {Quantum Christoffel Nonlinear Magnetization},
  author    = {Qiang, Xiao-Bin and Liu, Xiaoxiong and Lu, Hai-Zhou and Xie, X. C.},
  journal   = {Phys. Rev. Lett.},
  volume    = {136},
  issue     = {5},
  pages     = {056302},
  numpages  = {10},
  year      = {2026},
  month     = {Feb},
  publisher = {American Physical Society},
  doi       = {10.1103/4kmy-59l9},
  url       = {https://link.aps.org/doi/10.1103/4kmy-59l9}
}

@article{Wang2026SMOKE,
  title     = {Probing quantum geometric nonlinear magnetization via second-harmonic magneto-optical Kerr effect},
  author    = {Qian, Xuan and Qiang, Xiao-Bin and Zhu, Wenkai and Huang, Yuqing and Chen, Yiyuan and Lu, Hai-Zhou and Ji, Yang and Wang, Kaiyou},
  journal   = {Phys. Rev. B},
  volume    = {113},
  issue     = {4},
  pages     = {L041407},
  numpages  = {7},
  year      = {2026},
  month     = {Jan},
  publisher = {American Physical Society},
  doi       = {10.1103/dww3-vm74},
  url       = {https://link.aps.org/doi/10.1103/dww3-vm74}
}

@article{Liao2025nonlinear-OM,
  title     = {Current-induced anomalous Hall effect and nonlinear magnetoelectric effect in the Weyl semimetal $\mathrm{WT}{\mathrm{e}}_{2}$},
  author    = {Ye, Xing-Guo and Zhu, Peng-Fei and Xu, Wen-Zheng and Wang, An-Qi and Liao, Zhi-Min},
  journal   = {Phys. Rev. B},
  volume    = {111},
  issue     = {8},
  pages     = {085430},
  numpages  = {6},
  year      = {2025},
  month     = {Feb},
  publisher = {American Physical Society},
  doi       = {10.1103/PhysRevB.111.085430},
  url       = {https://link.aps.org/doi/10.1103/PhysRevB.111.085430}
}

@article{Liao2024nonlinear-OM,
  title     = {Nonlinear spin and orbital Edelstein effect in $\mathrm{WT}{\mathrm{e}}_{2}$},
  author    = {Ye, Xing-Guo and Zhu, Peng-Fei and Xu, Wen-Zheng and Zhao, Tong-Yang and Liao, Zhi-Min},
  journal   = {Phys. Rev. B},
  volume    = {110},
  issue     = {20},
  pages     = {L201407},
  numpages  = {7},
  year      = {2024},
  month     = {Nov},
  publisher = {American Physical Society},
  doi       = {10.1103/PhysRevB.110.L201407},
  url       = {https://link.aps.org/doi/10.1103/PhysRevB.110.L201407}
}

@article{Feng2025Quantum,
  author           = {Xukun Feng and Weikang Wu and Hui Wang and Weibo Gao and Lay Kee Ang and Y.X. Zhao and Cong Xiao and Shengyuan A. Yang},
  journal          = {Materials Today Quantum},
  title            = {Quantum metric nonlinear spin-orbit torque enhanced by topological bands},
  year             = {2025},
  issn             = {2950-2578},
  pages            = {100040},
  volume           = {6},
  doi              = {https://doi.org/10.1016/j.mtquan.2025.100040},
  modificationdate = {2025-05-21T13:46:38},
  url              = {https://www.sciencedirect.com/science/article/pii/S2950257825000186}
}

@article{lee2026Orbital,
  title   = {Orbital Exchange-Mediated Current Control of Magnetism},
  author  = {Lee, Geun-Hee and Kim, Kyoung-Whan and Lee, Kyung-Jin},
  year    = 2026,
  journal = {Nat. Commun.},
  volume  = {17},
  number  = {1},
  pages   = {2236},
  issn    = {2041-1723},
  doi     = {10.1038/s41467-026-68846-x},
  url     = {https://www.nature.com/articles/s41467-026-68846-x},
  langid  = {english}
}

@article{Klaui2026orbital,
  title   = {Orbital Magnetoresistance in the Antiferromagnet CoO Driven by Dynamic Orbital Angular Momentum},
  author  = {Schmitt, Christin and Krishnia, Sachin and Zeer, Mahmoud and {Gal{\'i}ndez-Ruales}, Edgar and Loyal, Mehak and K{\"o}hler, Jonas and Micus, Luca and Kikkawa, Takashi and Arisawa, Hiroki and Denneulin, Thibaud and Kov{\'a}cs, Andr{\'a}s and Xu, Renyou and Tran, Duc and Kronast, Florian and Go, Dongwook and Pourovskii, Leonid V. and {Dunin-Borkowski}, Rafal E. and Kuschel, Timo and Le{\v z}ai{\'c}, Marjana and Sinova, Jairo and Saitoh, Eiji and Jakob, Gerhard and Gomonay, Olena and Mokrousov, Yuriy and Kl{\"a}ui, Mathias},
  year    = 2026,
  journal = {Science},
  volume  = {393},
  number  = {6806},
  pages   = {76--79},
  issn    = {0036-8075, 1095-9203},
  doi     = {10.1126/science.adw1808},
  url     = {https://www.science.org/doi/10.1126/science.adw1808},
  langid  = {english}
}

@article{Awschalom2009,
  title     = {Spintronics without magnetism},
  author    = {Awschalom, David and Samarth, Nitin},
  journal   = {Physics},
  volume    = {2},
  pages     = {50},
  year      = {2009},
  publisher = {APS},
  doi       = {10.1103/Physics.2.50},
  url       = {http://link.aps.org/doi/10.1103/Physics.2.50}
}

@article{furukawa2017observation,
  title     = {Observation of current-induced bulk magnetization in elemental tellurium},
  author    = {Furukawa, Tetsuya and Shimokawa, Yuri and Kobayashi, Kaya and Itou, Tetsuaki},
  journal   = {Nat. Commun.},
  volume    = {8},
  number    = {1},
  pages     = {954},
  year      = {2017},
  publisher = {Nature Publishing Group UK London},
  doi       = {10.1038/s41467-017-01093-3},
  url       = {https://doi.org/10.1038/s41467-017-01093-3}
}

@article{Murakami2015,
  title     = {Current-induced orbital and spin magnetizations in crystals with helical structure},
  author    = {Yoda, Taiki and Yokoyama, Takehito and Murakami, Shuichi},
  journal   = {Sci. Rep.},
  volume    = {5},
  number    = {1},
  pages     = {12024},
  year      = {2015},
  publisher = {Nature Publishing Group UK London},
  doi       = {10.1038/srep12024},
  url       = {https://doi.org/10.1038/srep12024}
}

@article{Mak2017valley,
  title     = {Valley magnetoelectricity in single-layer MoS2},
  author    = {Lee, Jieun and Wang, Zefang and Xie, Hongchao and Mak, Kin Fai and Shan, Jie},
  journal   = {Nat. Mater.},
  volume    = {16},
  number    = {9},
  pages     = {887--891},
  year      = {2017},
  publisher = {Nature Publishing Group UK London},
  doi       = {10.1038/nmat4931},
  url       = {https://doi.org/10.1038/nmat4931}
}

@article{Murakami2018orbital,
  title     = {Orbital Edelstein effect as a condensed-matter analog of solenoids},
  author    = {Yoda, Taiki and Yokoyama, Takehito and Murakami, Shuichi},
  journal   = {Nano Lett.},
  volume    = {18},
  number    = {2},
  pages     = {916--920},
  year      = {2018},
  publisher = {ACS Publications},
  doi       = {10.1021/acs.nanolett.7b04300},
  url       = {https://doi.org/10.1021/acs.nanolett.7b04300}
}

@article{peng_High_2019,
  title     = {High {Phase} {Purity} of {Large}-{Sized} {1T}'-{MoS2} {Monolayers} with {2D} {Superconductivity}},
  volume    = {31},
  copyright = {© 2019 WILEY-VCH Verlag GmbH \& Co. KGaA, Weinheim},
  issn      = {1521-4095},
  url       = {https://onlinelibrary.wiley.com/doi/abs/10.1002/adma.201900568},
  doi       = {10.1002/adma.201900568},
  number    = {19},
  urldate   = {2024-03-24},
  journal   = {Adv. Mater.},
  author    = {Peng, Jing and Liu, Yuhua and Luo, Xiao and Wu, Jiajing and Lin, Yue and Guo, Yuqiao and Zhao, Jiyin and Wu, Xiaojun and Wu, Changzheng and Xie, Yi},
  year      = {2019},
  pages     = {1900568}
}

@article{Stern2006,
  title     = {Current-Induced Polarization and the Spin Hall Effect at Room Temperature},
  author    = {Stern, N. P. and Ghosh, S. and Xiang, G. and Zhu, M. and Samarth, N. and Awschalom, D. D.},
  journal   = {Phys. Rev. Lett.},
  volume    = {97},
  issue     = {12},
  pages     = {126603},
  numpages  = {4},
  year      = {2006},
  month     = {Sep},
  publisher = {American Physical Society},
  doi       = {10.1103/PhysRevLett.97.126603},
  url       = {https://link.aps.org/doi/10.1103/PhysRevLett.97.126603}
}

@article{Zhong2016,
  title     = {Gyrotropic Magnetic Effect and the Magnetic Moment on the Fermi Surface},
  author    = {Zhong, Shudan and Moore, Joel E. and Souza, Ivo},
  journal   = {Phys. Rev. Lett.},
  volume    = {116},
  issue     = {7},
  pages     = {077201},
  numpages  = {6},
  year      = {2016},
  month     = {Feb},
  publisher = {American Physical Society},
  doi       = {10.1103/PhysRevLett.116.077201},
  url       = {https://link.aps.org/doi/10.1103/PhysRevLett.116.077201}
}

@article{Wang2024IPHE-exp,
  title     = {Orbital Magneto-Nonlinear Anomalous Hall Effect in Kagome Magnet ${\mathrm{Fe}}_{3}{\mathrm{Sn}}_{2}$},
  author    = {Wang, Lujunyu and Zhu, Jiaojiao and Chen, Haiyun and Wang, Hui and Liu, Jinjin and Huang, Yue-Xin and Jiang, Bingyan and Zhao, Jiaji and Shi, Hengjie and Tian, Guang and Wang, Haoyu and Yao, Yugui and Yu, Dapeng and Wang, Zhiwei and Xiao, Cong and Yang, Shengyuan A. and Wu, Xiaosong},
  journal   = {Phys. Rev. Lett.},
  volume    = {132},
  issue     = {10},
  pages     = {106601},
  numpages  = {7},
  year      = {2024},
  month     = {Mar},
  publisher = {American Physical Society},
  doi       = {10.1103/PhysRevLett.132.106601},
  url       = {https://link.aps.org/doi/10.1103/PhysRevLett.132.106601}
}

@article{Wang2024IPHE,
  title     = {Orbital Origin of the Intrinsic Planar Hall Effect},
  author    = {Wang, Hui and Huang, Yue-Xin and Liu, Huiying and Feng, Xiaolong and Zhu, Jiaojiao and Wu, Weikang and Xiao, Cong and Yang, Shengyuan A.},
  journal   = {Phys. Rev. Lett.},
  volume    = {132},
  issue     = {5},
  pages     = {056301},
  numpages  = {6},
  year      = {2024},
  month     = {Jan},
  publisher = {American Physical Society},
  doi       = {10.1103/PhysRevLett.132.056301},
  url       = {https://link.aps.org/doi/10.1103/PhysRevLett.132.056301}
}

@article{Guo2024,
  title     = {Extrinsic contribution to nonlinear current induced spin polarization},
  author    = {Guo, Ruda and Huang, Yue-Xin and Yang, Xiaoxin and Liu, Yi and Xiao, Cong and Yuan, Zhe},
  journal   = {Phys. Rev. B},
  volume    = {109},
  issue     = {23},
  pages     = {235413},
  numpages  = {10},
  year      = {2024},
  month     = {Jun},
  publisher = {American Physical Society},
  doi       = {10.1103/PhysRevB.109.235413},
  url       = {https://link.aps.org/doi/10.1103/PhysRevB.109.235413}
}

@article{Han2024room,
  title     = {Room-Temperature Flexible Manipulation of the Quantum-Metric Structure in a Topological Chiral Antiferromagnet},
  author    = {Han, Jiahao and Uchimura, Tomohiro and Araki, Yasufumi and Yoon, Ju-Young and Takeuchi, Yutaro and Yamane, Yuta and Kanai, Shun and Ieda, Jun'ichi and Ohno, Hideo and Fukami, Shunsuke},
  year      = 2024,
  journal   = {Nat. Phys.},
  volume    = {20},
  number    = {7},
  pages     = {1110--1117},
  publisher = {Nature Publishing Group},
  issn      = {1745-2481},
  doi       = {10.1038/s41567-024-02476-2},
  url       = {https://www.nature.com/articles/s41567-024-02476-2},
  langid    = {english}
}

@article{Gao2023QM,
  author           = {Gao, Anyuan and Liu, Yu-Fei and Qiu, Jian-Xiang and Ghosh, Barun and V. Trevisan, Thaís and Onishi, Yugo and Hu, Chaowei and Qian, Tiema and Tien, Hung-Ju and Chen, Shao-Wen and Huang, Mengqi and Bérubé, Damien and Li, Houchen and Tzschaschel, Christian and Dinh, Thao and Sun, Zhe and Ho, Sheng-Chin and Lien, Shang-Wei and Singh, Bahadur and Watanabe, Kenji and Taniguchi, Takashi and Bell, David C. and Lin, Hsin and Chang, Tay-Rong and Du, Chunhui Rita and Bansil, Arun and Fu, Liang and Ni, Ni and Orth, Peter P. and Ma, Qiong and Xu, Su-Yang},
  journal          = {Science},
  title            = {Quantum metric nonlinear Hall effect in a topological antiferromagnetic heterostructure},
  year             = {2023},
  issn             = {1095-9203},
  month            = jul,
  number           = {6654},
  pages            = {181--186},
  volume           = {381},
  creationdate     = {2024-01-09T21:17:14},
  doi              = {10.1126/science.adf1506},
  modificationdate = {2024-01-09T21:17:30},
  publisher        = {American Association for the Advancement of Science (AAAS)}
}

@article{Xiao2022NLSOT,
  title     = {Intrinsic Nonlinear Electric Spin Generation in Centrosymmetric Magnets},
  author    = {Xiao, Cong and Liu, Huiying and Wu, Weikang and Wang, Hui and Niu, Qian and Yang, Shengyuan A.},
  journal   = {Phys. Rev. Lett.},
  volume    = {129},
  issue     = {8},
  pages     = {086602},
  numpages  = {6},
  year      = {2022},
  month     = {Aug},
  publisher = {American Physical Society},
  doi       = {10.1103/PhysRevLett.129.086602},
  url       = {https://link.aps.org/doi/10.1103/PhysRevLett.129.086602}
}

@article{Wang2022,
  title     = {Nonlinear antidamping spin-orbit torque originating from intraband transport on the warped surface of a topological insulator},
  author    = {Zhou, Yong-Long and Duan, Hou-Jian and Wu, Yong-jia and Deng, Ming-Xun and Wang, Lan and Culcer, Dimitrie and Wang, Rui-Qiang},
  journal   = {Phys. Rev. B},
  volume    = {105},
  issue     = {7},
  pages     = {075415},
  numpages  = {10},
  year      = {2022},
  month     = {Feb},
  publisher = {American Physical Society},
  doi       = {10.1103/PhysRevB.105.075415},
  url       = {https://link.aps.org/doi/10.1103/PhysRevB.105.075415}
}

@article{Kodama2024,
  title     = {Direct observation of current-induced nonlinear spin torque in Pt-Py bilayers},
  author    = {Kodama, Toshiyuki and Kikuchi, Nobuaki and Chiba, Takahiro and Okamoto, Satoshi and Ohno, Seigo and Tomita, Satoshi},
  journal   = {Phys. Rev. B},
  volume    = {109},
  issue     = {21},
  pages     = {214419},
  numpages  = {7},
  year      = {2024},
  month     = {Jun},
  publisher = {American Physical Society},
  doi       = {10.1103/PhysRevB.109.214419},
  url       = {https://link.aps.org/doi/10.1103/PhysRevB.109.214419}
}

@article{Gao2014,
  author    = {Gao, Yang and Yang, Shengyuan A. and Niu, Qian},
  journal   = {Phys. Rev. Lett.},
  title     = {Field Induced Positional Shift of Bloch Electrons and Its Dynamical Implications},
  year      = {2014},
  month     = {Apr},
  pages     = {166601},
  volume    = {112},
  issue     = {16},
  numpages  = {5},
  publisher = {American Physical Society},
  doi       = {10.1103/PhysRevLett.112.166601},
  url       = {https://link.aps.org/doi/10.1103/PhysRevLett.112.166601}
}

@article{liu2022third,
  title     = {Berry connection polarizability tensor and third-order Hall effect},
  author    = {Liu, Huiying and Zhao, Jianzhou and Huang, Yue-Xin and Feng, Xiaolong and Xiao, Cong and Wu, Weikang and Lai, Shen and Gao, Wei-bo and Yang, Shengyuan A.},
  journal   = {Phys. Rev. B},
  volume    = {105},
  issue     = {4},
  pages     = {045118},
  numpages  = {7},
  year      = {2022},
  month     = {Jan},
  publisher = {American Physical Society},
  doi       = {10.1103/PhysRevB.105.045118},
  url       = {https://link.aps.org/doi/10.1103/PhysRevB.105.045118}
}

@article{xiao2023,
  title     = {Time-Reversal-Even Nonlinear Current Induced Spin Polarization},
  author    = {Xiao, Cong and Wu, Weikang and Wang, Hui and Huang, Yue-Xin and Feng, Xiaolong and Liu, Huiying and Guo, Guang-Yu and Niu, Qian and Yang, Shengyuan A.},
  journal   = {Phys. Rev. Lett.},
  volume    = {130},
  issue     = {16},
  pages     = {166302},
  numpages  = {6},
  year      = {2023},
  month     = {Apr},
  publisher = {American Physical Society},
  doi       = {10.1103/PhysRevLett.130.166302},
  url       = {https://link.aps.org/doi/10.1103/PhysRevLett.130.166302}
}

@article{jungwirth2002,
  title     = {Anomalous Hall Effect in Ferromagnetic Semiconductors},
  author    = {Jungwirth, T. and Niu, Qian and MacDonald, A. H.},
  journal   = {Phys. Rev. Lett.},
  volume    = {88},
  issue     = {20},
  pages     = {207208},
  year      = {2002},
  publisher = {American Physical Society},
  doi       = {10.1103/PhysRevLett.88.207208},
  url       = {https://link.aps.org/doi/10.1103/PhysRevLett.88.207208}
}

@article{qian2014,
  title   = {Quantum spin Hall effect in two-dimensional transition
             metal dichalcogenides},
  author  = {Qian, Xiaofeng and Liu, Junwei and Fu, Liang and Li, Ju},
  journal = {Science},
  volume  = {346},
  number  = {6215},
  pages   = {1344--1347},
  year    = {2014},
  doi     = {10.1126/science.1256815},
  url     = {https://doi.org/10.1126/science.1256815}
}

@article{tang2018,
  title   = {Electronic structure of monolayer 1T'-MoTe2 grown by
             molecular beam epitaxy},
  author  = {Tang, Shujie and Zhang, Chaofan and Jia, Chunjing and Ryu,
             Hyejin and Hwang, Choongyu and Hashimoto, Makoto and Lu,
             Donghui and Liu, Zhi and Devereaux, Thomas P and Shen,
             Zhi-Xun and others},
  journal = {APL Mater.},
  volume  = {6},
  number  = {2},
  pages   = {026601},
  year    = {2018},
  doi     = {10.1021/acs.nanolett.6b01342},
  url     = {https://doi.org/10.1021/acs.nanolett.6b01342}
}

@article{naylor2016,
  title     = {Monolayer single-crystal 1T'-MoTe2 grown by chemical
               vapor deposition exhibits weak antilocalization effect},
  author    = {Naylor, Carl H and Parkin, William M and Ping, Jinglei and
               Gao, Zhaoli and Zhou, Yu Ren and Kim, Youngkuk and
               Streller, Frank and Carpick, Robert W and Rappe, Andrew M
               and Drndic, Marija and others},
  journal   = {Nano Lett.},
  volume    = {16},
  number    = {7},
  pages     = {4297--4304},
  year      = {2016},
  publisher = {ACS Publications},
  doi       = {10.1063/1.5004700},
  url       = {https://doi.org/10.1063/1.5004700}
}

@article{pikus1978,
  title   = {New photogalvanic effect in gyrotropic crystals},
  author  = {Ivchenko, E. L. and Pikus, G. E.},
  journal = {JETP Lett.},
  volume  = {27},
  pages   = {604},
  year    = {1978}
}

@article{aronov1989,
  title   = {Nuclear electric resonance and orientation of carrier
             spins by an electric field},
  author  = {Aronov, A. G. and Lyanda-Geller, Y.},
  journal = {JETP Lett.},
  volume  = {50},
  pages   = {431},
  year    = {1989}
}

@article{jungwirth2018,
  author  = {Olejn{\'\i}k, Kamil and Seifert, Tom and Ka{\v{s}}par,
             Zden{\v{e}}k and Nov{\'a}k, V{\'\i}t and Wadley, Peter and
             Campion, Richard P and Baumgartner, Manuel and Gambardella,
             Pietro and N{\v{e}}mec, Petr and Wunderlich, Joerg and
             Sinova, Jairo and Ku{\v{z}}el, Petr and M{\"u}ller, Melanie
             and Kampfrath, Tobias and Jungwirth, Tomas},
  title   = {Terahertz electrical writing speed in an antiferromagnetic
             memory},
  journal = {Sci. Adv.},
  volume  = {4},
  number  = {3},
  pages   = {eaar3566},
  year    = {2018},
  doi     = {10.1126/sciadv.aar3566},
  url     = {https://www.science.org/doi/abs/10.1126/sciadv.aar3566}
}

@article{xiao2010,
  author    = {Xiao, Di and Chang, Ming-Che and Niu, Qian},
  journal   = {Rev. Mod. Phys.},
  title     = {Berry phase effects on electronic properties},
  year      = {2010},
  month     = {Jul},
  pages     = {1959--2007},
  volume    = {82},
  issue     = {3},
  publisher = {American Physical Society},
  doi       = {10.1103/RevModPhys.82.1959},
  url       = {https://link.aps.org/doi/10.1103/RevModPhys.82.1959}
}

@article{gao2015,
  author    = {Gao, Yang and Yang, Shengyuan A. and Niu, Qian},
  journal   = {Phys. Rev. B},
  title     = {Geometrical effects in orbital magnetic susceptibility},
  year      = {2015},
  month     = {Jun},
  pages     = {214405},
  volume    = {91},
  issue     = {21},
  numpages  = {12},
  publisher = {American Physical Society},
  doi       = {10.1103/PhysRevB.91.214405},
  url       = {https://link.aps.org/doi/10.1103/PhysRevB.91.214405}
}

@article{edelstein,
  title   = {Spin polarization of conduction electrons induced by
             electric current in two-dimensional asymmetric electron
             systems},
  author  = {Edelstein, Y. M.},
  journal = {Solid State Commun.},
  volume  = {73},
  issue   = {3},
  pages   = {233},
  year    = {1990},
  doi     = {10.1016/0038-1098(90)90963-C},
  url     = {https://doi.org/10.1016/0038-1098(90)90963-C}
}

@article{kato2004,
  title     = {Current-Induced Spin Polarization in Strained
               Semiconductors},
  author    = {Kato, Y. K. and Myers, R. C. and Gossard, A. C. and
               Awschalom, D. D.},
  journal   = {Phys. Rev. Lett.},
  volume    = {93},
  issue     = {17},
  pages     = {176601},
  numpages  = {4},
  year      = {2004},
  month     = {Oct},
  publisher = {American Physical Society},
  doi       = {10.1103/PhysRevLett.93.176601},
  url       = {https://link.aps.org/doi/10.1103/PhysRevLett.93.176601}
}

@article{culcer2007,
  title     = {Generation of Spin Currents and Spin Densities in Systems
               with Reduced Symmetry},
  author    = {Culcer, Dimitrie and Winkler, R.},
  journal   = {Phys. Rev. Lett.},
  volume    = {99},
  issue     = {22},
  pages     = {226601},
  numpages  = {4},
  year      = {2007},
  month     = {Nov},
  publisher = {American Physical Society},
  doi       = {10.1103/PhysRevLett.99.226601},
  url       = {https://link.aps.org/doi/10.1103/PhysRevLett.99.226601}
}

@article{wang2021,
  title     = {Intrinsic Nonlinear Hall Effect in Antiferromagnetic
               Tetragonal CuMnAs},
  author    = {Wang, Chong and Gao, Yang and Xiao, Di},
  journal   = {Phys. Rev. Lett.},
  volume    = {127},
  issue     = {27},
  pages     = {277201},
  numpages  = {6},
  year      = {2021},
  month     = {Dec},
  publisher = {American Physical Society},
  doi       = {10.1103/PhysRevLett.127.277201},
  url       = {https://link.aps.org/doi/10.1103/PhysRevLett.127.277201}
}

@article{liu2021,
  title     = {Intrinsic Second-Order Anomalous Hall Effect and Its
               Application in Compensated Antiferromagnets},
  author    = {Liu, Huiying and Zhao, Jianzhou and Huang, Yue-Xin and Wu,
               Weikang and Sheng, Xian-Lei and Xiao, Cong and Yang,
               Shengyuan A.},
  journal   = {Phys. Rev. Lett.},
  volume    = {127},
  issue     = {27},
  pages     = {277202},
  numpages  = {6},
  year      = {2021},
  month     = {Dec},
  publisher = {American Physical Society},
  doi       = {10.1103/PhysRevLett.127.277202},
  url       = {https://link.aps.org/doi/10.1103/PhysRevLett.127.277202}
}

@article{ma2019,
  title   = {Observation of the nonlinear Hall effect under
             time-reversal-symmetric conditions},
  author  = {Ma, Qiong and Xu, Su-Yang and Shen, Huitao and MacNeill,
             David and Fatemi, Valla and Chang, Tay-Rong and Mier
             Valdivia, Andr\'es M. and Wu, Sanfeng and Du, Zongzheng and
             Hsu, Chuang-Han and Fang, Shiang and Gibson, Quinn D. and
             Watanabe, Kenji and Taniguchi, Takashi and Cava, Robert J.
             and Kaxiras, Efthimios and Lu, Hai-Zhou and Lin, Hsin and
             Fu, Liang and Gedik, Nuh and Jarillo-Herrero, Pablo},
  journal = {Nature},
  volume  = {565},
  pages   = {337},
  year    = {2019},
  doi     = {10.1038/s41586-018-0807-6},
  url     = {https://doi.org/10.1038/s41586-018-0807-6}
}

@article{sundaram_wave-packet_1999,
  title      = {Wave-packet dynamics in slowly perturbed crystals:
                {Gradient} corrections and {Berry}-phase effects},
  volume     = {59},
  issn       = {0163-1829, 1095-3795},
  shorttitle = {Wave-packet dynamics in slowly perturbed crystals},
  url        = {https://link.aps.org/doi/10.1103/PhysRevB.59.14915},
  doi        = {10.1103/PhysRevB.59.14915},
  number     = {23},
  urldate    = {2018-09-13},
  journal    = {Phys. Rev. B},
  author     = {Sundaram, Ganesh and Niu, Qian},
  month      = jun,
  year       = {1999},
  pages      = {14915--14925}
}

@article{chang_berry_1995,
  title   = {Berry {Phase}, {Hyperorbits}, and the {Hofstadter}
             {Spectrum}},
  volume  = {75},
  issn    = {0031-9007, 1079-7114},
  url     = {https://link.aps.org/doi/10.1103/PhysRevLett.75.1348},
  doi     = {10.1103/PhysRevLett.75.1348},
  number  = {7},
  urldate = {2022-08-05},
  journal = {Phys. Rev. Lett.},
  author  = {Chang, Ming-Che and Niu, Qian},
  month   = aug,
  year    = {1995},
  pages   = {1348--1351}
}

\end{document}